\documentclass[amsmath,amssymb,twocolumn]{revtex4}
\usepackage{subfig}
\usepackage{graphicx}
\usepackage{dcolumn}
\usepackage{bm}
\usepackage{amsmath,mathrsfs}
\usepackage{graphicx,epsfig}
\usepackage{hyperref}
\usepackage{epsf}
\usepackage{color}

\begin{document}


\title{{\bf Non-exotic wormholes in $f(R,L_m)$ gravity}}

\author{S. Rastgoo}\email{rastgoo@sirjantech.ac.ir}
\author{ F. Parsaei }\email{fparsaei@gmail.com}

\affiliation{ Physics Department , Sirjan University of Technology, Sirjan 78137, Iran.}

\date{\today}


\begin{abstract}
\par In the present analysis, we examine the potential existence of generalized wormhole models within the framework of newly developed extended $f(R,L_m)$ gravity. We investigate both a linear model,  $f(R,L_m)=\alpha R+\beta L_m$, and a  non-linear model,  $f(R,L_m)=\frac{R}{2}+ L^\alpha_m$, to analyze traversable wormholes. By employing the variational approach, we derive modified versions of the field equations under the influence of an anisotropic matter source. A power-law shape function is applied, resulting in a linear equation of state for both radial and lateral pressures. Furthermore, we explore solutions characterized by a variable equation of state parameter. It was observed that the violation of energy conditions is influenced by the parameters $\alpha$ and $\beta$. A wide range of non-exotic wormhole solutions was discovered, dependent on the specific parameters of the model. We demonstrate that wormholes with power-law shape functions yield solutions that comply with the energy conditions in both linear and non-linear forms of $f(R, L_m)$. It is shown that the non-exotic wormhole solutions obtained within this framework are not isotropic.
  \\
\end{abstract}

\maketitle
\section{Introduction}
The mysterious occurrence of wormholes, which are theoretical entities that connect far-flung areas of spacetime, has fascinated both physicists and the general populace. These speculative formations, frequently illustrated as cosmic shortcuts linking distinct locations in the Universe, have undergone thorough examination within the context of general relativity (GR). Historically, Flamm \cite{flamm} was the first to introduce the idea of a wormhole. The notion of "bridges" was proposed in 1935 by physicists Einstein and Rosen \cite{Rosen}, which is well-known as the Einstein–Rosen bridge. It was believed to be non-traversable, as nothing, not even photons, could pass through its throat due to its swift expansion and contraction \cite{Rosen}. The phrase 'wormhole' was first introduced in \cite{wheeler} by Misner and Wheeler.
In 1988, Morris and Thorne \cite{WH} introduced the concept of static traversable wormholes. Morris and Thorne investigated the concept of developing wormholes that could potentially be traversable by matter and possibly facilitate time travel, which led to the initial establishment of the idea of traversable wormholes \cite{WH}. In the context of GR, the presence of exotic matter is essential for maintaining the stability of a wormhole's throat \cite{Visser}. In classical GR, solutions involving wormholes contravene all energy conditions (ECs). Exotic matter produces a repulsive force that stops the wormhole from collapsing.
Numerous efforts have been made to reduce the utilization of exotic matter. These methods may reduce the necessity for exotic matter to satisfy the conditions of spacetime geometry. In these theoretical frameworks, exotic matter is confined to a designated area of spacetime. Prominent examples of these are thin-shell wormholes
\cite{cut, cut1, cut2, cut3}, those characterized by a variable equation of state (EoS) \cite{Remo, variable}, and wormholes that are defined by a polynomial EoS \cite{foad}.

Following the discovery of the Universe's accelerated expansion, researchers have put forward a variety of concepts that can be classified into two separate categories of modified propositions: those pertaining to modified matter and those associated with modified curvature. Phantom fluid, classified as a particular form of dark energy (DE), may be regarded as a type of exotic matter. So, extensive research on phantom wormholes that contravene ECs has been conducted in the existing literature \cite{phantom, phantom2, phantom1}.

As the second method, alterations to the conventional theory of gravity, specifically GR, have primarily been driven by the shortcomings of GR in explaining the accelerated expansion of the Universe at late times. Also, the quest to eliminate the need for exotic matter or to alleviate ECs has propelled the exploration of wormholes in alternative gravitational theories. In this sense, wormholes are studied in Braneworld \cite{b, b1, b2, b3}, Born-Infeld theory \cite{Bo, Bo1}, quadratic gravity \cite{quad, quad1}, Einstein-Cartan gravity \cite{Cartan, Cartan1, Cartan2}, Rastall gravity \cite{Rast}, Rastall–Rainbow gravity \cite{RaR, RaR1},  $f(Q)$ gravity \cite{fq, fq2, fq3, fq4, fq44, fq5, fq6}, $f(R)$ gravity \cite{Nojiri, fR0, fR11, fR22, fR33, fR44,fR55}, Ricci inverse gravity \cite{inverse}, $f(T,\mathcal{T})$ gravity \cite{Must1, Err, Riz, Must2, foad3}, and $f(R,T)$ gravity \cite{Azizi, Moa, Zub, Shar, Rosa, fr2, fr3, Sha, Ban, Sarkar, SR1, foad4}. Due to the modification of Einstein field equations in these scenarios, some of them are capable to elaborate the problem of exotic matter in the wormhole theory.

The generalization of GR to $f(R)$ gravity has been proposed by Harko \cite{Ha}. Another revised theory of gravity expands upon the classical $f(R)$ gravity by incorporating a dependence on a Lagrangian density ($L_m$), which is associated with the matter content of the Universe, in addition to the Ricci scalar ($R$). The $f(R, L_m)$ represents the outcomes of the direct interaction between the scalar curvature and the matter Lagrangian \cite{Ha}. This has led to the derivation of the mass-particle equation of motion, revealing an additional force that acts perpendicular to the four-velocity. The $f (R, L_m) $  modified gravity leads to the non-vanishing covariant divergence of the energy-momentum tensor, resulting in additional forces that cause test particles to follow non-geodesic paths. Moreover, these models do not comply with the equivalence principle, which has been constrained by solar system experiments. This raises important considerations for the validity and applicability of $f (R, L_m) $  gravity in explaining gravitational phenomena. The cosmological models that incorporate curvature–matter couplings possess considerable astrophysical and cosmological applications. Numerous scholars have examined cosmology through the lens of the $f (R, L_m)$ gravity framework \cite{ap, ap0, PLA, ap1, ap2, ap3}.
Recently, a significant number of papers focusing on the topic of wormholes within the context of $f (R, L_m)$ gravity have been published \cite{L1, L2, L3, L33, L4, L5, L6, L7, L8, L9, L10, L11, L12}. In \cite{L1}, three distinct models of shape functions have been examined within the framework of $f(R, L_m)$ that contravene the ECs.
 Jaybhaye et al. have investigated traversable wormhole solutions within the framework of the $f (R, L_m)$ theory of gravity by utilizing DM halo profiles \cite{L11}.
Soni et al. have examined traversable wormhole solutions in conjunction with a charge within the $f (R, L_m)$ gravity framework \cite{L33}. They have analyzed the effect of charge on wormholes through two distinct approaches for a constant redshift function. In \cite{L3}, solutions involving wormholes in the context of curvature–matter coupling gravity, which are underpinned by non-commutative geometry and conformal symmetry, are examined. In \cite{L4}, the Karmarkar condition has been employed to ascertain the geometry of the wormhole within the context of $f(R, L_m)$ gravity. The traversable wormholes, along with the GUP corrected Casimir effect in $f(R, L_m)$ gravity, are examined in \cite{L7, L12}. Additionally, the Casimir wormhole solutions within the framework of $f(R, L_m)$ gravity are analyzed in \cite{L9}.
 Naser and colleagues have explored the consequences of $f (R, L_m)$ theory regarding the existence of charged wormhole solutions \cite{L2}. They have acquired the wormhole solutions by considering three scenarios: a linear barotropic EoS, an anisotropic EoS, and an isotropic EoS. The influence of the Einasto spike and the isothermal DM density profile on the geometry of a wormhole in $f (R, L_m)$  gravity has been studied in \cite{L8}. In \cite{L10}, the authors employed an anisotropic matter source along with a specific form of energy density to illustrate cold DM halo and quantum wave DM halo, which were utilized to derive two distinct wormhole solutions.

To the best of our understanding, all suggested wormhole solutions within the framework of $f (R, L_m)$ gravity violate the classical ECs. Consequently, the main aim of this paper is to discover wormhole solutions that adhere to these ECs.
To advance our understanding of
$f (R, L_m)$ gravity, we begin by deriving the field equations within this framework. Our goal is to establish a clear connection between the field equations of GR and those formulated in the context of
$f (R, L_m)$ gravity. Subsequently, we utilize both linear and nonlinear functions for $f (R, L_m)$ to investigate solutions that fulfill the ECs.

The structure of our paper is outlined as follows: In Sec.\ref{sec2}, we introduce the fundamental concept of wormholes. Following this, we deliver a brief summary of the $f (R, L_m)$ theory alongside the classical ECs. In Sec.\ref{sec3}, we examine two functions for $f (R, L_m)$ and subsequently utilize the relevant field equations to propose solutions that adhere to the ECs. This Section also includes an examination of the physical characteristics associated with these solutions.  In the last Section, we summarize our results.  Throughout this manuscript, we have utilized gravitational units by setting $c = 8 \pi G = 1$.

\section{Basic formulation of wormhole and $f(R,L_m)$ gravity} \label{sec2}
The general line element of the Morris-Thorne wormhole, characterized by spherical symmetry and static properties, is articulated as follows
\begin{equation}\label{1}
ds^2=-U(r)dt^2+\frac{dr^2}{1-\frac{b(r)}{r}}+r^2(d\theta^2+\sin^2\theta,
d\phi^2)
\end{equation}
where $U(r)=\exp (2\phi(r))$.
Here, $b(r)$ is known as the shape function, which has a direct impact on the curvature and configuration of the throat region, as it represents the geometry of the wormhole. In this context, $\phi(r)$  represents the redshift function associated with the phenomenon of gravitational redshift.
The throat condition implies that
\begin{equation}\label{2}
b(r_0)=r_0
\end{equation}
where  $r_0$ is the wormhole throat. Additionally, two additional conditions need to be satisfied to ensure the presence of a traversable wormhole,
\begin{equation}\label{3}
b'(r_0)<1
\end{equation}
and
\begin{equation}\label{4}
b(r)<r,\ \ {\rm for} \ \ r>r_0.
\end{equation}
Equation (\ref{3}) is widely recognized as the flaring-out condition. This equation indicates a breach of the NEC within the framework of GR. To maintain consistency with the cosmos on a large scale, the metric elements must adhere to the asymptotically flat condition,
\begin{equation}\label{5}
\lim_{r\rightarrow \infty}\frac{b(r)}{r}=0,\qquad   \lim_{r\rightarrow \infty}U(r)=1.
\end{equation}
To guarantee that there is no event horizon, the redshift function  must stay finite at every point. To improve simplicity, we concentrate on solutions that include a constant redshift function. The vanishing redshift function exemplifies a basic and commonly utilized scenario in the examination of wormholes within the context of GR or its modified gravitational theories. The vanishing redshift function signifies the absence of tidal forces.
In this article, we consider an anisotropic fluid represented by the tensor $T^{\mu}_{\nu}=diag[-\rho, p, p_t, p_t]$. The energy density is represented by $\rho$, whereas $p$ and $p_t$ signify the radial and tangential pressures, respectively, both of which depend exclusively on the radial coordinate $r$.

In the subsequent Section, we will present a brief summary of the $f(R, L_m)$ gravity. The action in this scenario is expressed in the form
\begin{equation}\label{6a}
S= \int{f(R,L_m)\sqrt{-g}\text{d}^4x}\,
\end{equation}
where $R$ represents the scalar curvature, $L_m$ denotes the matter Lagrangian, and $f(R, L_m)$ is an arbitrary function of these two variables. The field equation derived from the variation of the general action \ref{6a} related to the metric tensor $g_{\mu \nu}$ is presented as follows:
 \begin{multline}\label{6b}
f_R R_{\mu\nu} + (g_{\mu\nu} \square - \nabla_\mu \nabla_\nu)f_R - \frac{1}{2} (f-f_{L_m}L_m)g_{\mu\nu}\\
=\frac{1}{2} f_{L_m} T_{\mu\nu}.
\end{multline}

In this context, $f_R\equiv \frac{\partial f}{\partial R}$, $f_{L_m}\equiv \frac{\partial f}{\partial L_m}$, and the energy–momentum tensor $L_m$ assume the resulting form
\begin{equation}\label{8}
T_{ij}= - \frac{2}{\sqrt{-g}}\left[\frac{\partial(\sqrt{-g}L_{m})}{\partial g^{ij}}\right].
\end{equation}
We have assumed that $L_m$ depends exclusively on the metric component and is unaffected by its derivatives.

By integrating the metric (\ref{1}) and the anisotropic fluid  into the field equations (\ref{6b}), we obtain the subsequent field equations
\begin{multline}\label{3e}
\left( 1-\frac{b}{r} \right) \left[ -\left \lbrace \frac{2}{r}- \frac{(rb'-b)}{2r(r-b)}  \right\rbrace F' - F'' \right] \\
+ \frac{1}{2} \left( f-L_m f_{L_m} \right) = \frac{1}{2} f_{L_m} \rho\,,
\end{multline}
\begin{multline}\label{3f}
\left( 1-\frac{b}{r} \right) \left[ \left\lbrace    \frac{(rb'-b)}{r^2(r-b)} \right\rbrace F + \left\lbrace \frac{2}{r}- \frac{(rb'-b)}{2r(r-b)} \right\rbrace F' \right] \\
 - \frac{1}{2} \left( f-L_m f_{L_m} \right) = \frac{1}{2} f_{L_m} p_r\,,
\end{multline}
\begin{multline}\label{3g}
\left( 1-\frac{b}{r} \right) \left[ \left\lbrace \frac{(rb'+b)}{2r^2(r-b)} \right\rbrace F+ \left\lbrace \frac{2}{r}
- \frac{(rb'-b)}{2r(r-b)}  \right\rbrace F' + F'' \right] \\ - \frac{1}{2} \left( f-L_m f_{L_m} \right) = \frac{1}{2} f_{L_m} p_t\,.
\end{multline}
The final form of field equations depends on the $f(R, L_m)$ and $L_m$. We assume $L_m=\rho$ and two different models for $f(R, L_m)$ in the next Section to find wormhole solutions.

The ECs are mathematical restrictions imposed on the solutions of the Einstein equation to eliminate the potential existence of non-physical solutions. To qualify as a physically valid solution to Einstein's equations, a spacetime geometry must satisfy the ECs. These conditions are crucial for investigating the physically feasible arrangements of matter. To sustain a positive stress-energy tensor when matter is present, these ECs provide effective methodologies. The ECs, which include the NEC, dominant energy condition (DEC), weak energy condition (WEC), and strong energy condition (SEC), are clearly defined to assist in reaching this objective,
\begin{eqnarray}\label{21}
\textbf{NEC}&:& \rho+p\geq 0,\quad \rho+p_t\geq 0 \\
\label{21a}
\textbf{WEC}&:& \rho\geq 0, \rho+p\geq 0,\quad \rho+p_t\geq 0, \\
\textbf{DEC}&:& \rho\geq 0, \rho-|p|\geq 0,\quad \rho-|p_t|\geq 0, \\
\textbf{SEC}&:& \rho+p\geq 0,\, \rho+p_t\geq 0,\rho+p+2p_t \geq 0. \label{21b}
\end{eqnarray}
The ECs fundamentally stem from the Raychaudhuri equation, which outlines the dynamics of congruences of curves in spacetime. These conditions impose essential restrictions on the energy-momentum tensor and have significant theoretical applications in cosmology, black hole physics, singularity theorems, and modified gravity theories, providing critical insights into the nature of matter and energy in the Universe. Utilizing these equations allows us to clarify the complex trajectories followed by celestial bodies. At this point, as referenced in \cite{fq}, we will explore the ECs in the subsequent sections of this paper by defining the functions,
\begin{eqnarray}\label{22}
 H(r)&=& \rho+p ,\, H_1(r)= \rho+p_t,\, H_2(r)= \rho-|p|, \nonumber \\
 H_3(r)&=&\rho-|p_t|,\, H_4(r)= \rho+p+2p_t .
\end{eqnarray}
Researchers have utilized a range of strategies to uncover asymptotically flat wormhole solutions within the context of GR. The predominant approach involves examining an EoS to find the shape function by using field equations. Conversely, certain researchers adopt shape functions defined by free parameters, which they then adjust in order to identify solutions that conform to both physical and mathematical requirements. In the following Section, we will implement these techniques to locate asymptotically flat wormhole solutions. For the sake of simplicity, we will assume the value of $r_0=1$ in the following discussions throughout this document.

\section{Non-exotic Wormhole solutions }\label{sec3}
In the previous Section, the fundamental concept of wormhole, the modified gravity represented by $f(R, L_m)$, and the associated field equations are succinctly discussed. In this Section, we will assume two different models for the function $f(R, L_m)$ to find asymptotically flat wormhole solutions. In the initial model, a linear relationship with respect to $L_m$ is taken into account in the $f(R, L_m)$ function, whereas in the subsequent model, the $f(R, L_m)$ is characterized as a nonlinear function of $l_m$. These two models have been widely examined in the existing literature.

\subsection{ Linear model}\label{subsec1}
As the first model, we consider a linear form for $f(R, L_m)$ as follows
\begin{equation}\label{9a}
    f(R, L_m)=\alpha \frac{R}{2}+\beta L_m.
\end{equation}
where $\alpha$ and $\beta$ are free parameters. This represents a straightforward linear adjustment in which $\beta$ serves as a coupling constant. It linearly alters the Einstein-Hilbert action in relation to the matter Lagrangian, facilitating a connection between gravity and matter. Using (\ref{9a}) in (\ref{3e}-\ref{3g}) gives

\begin{equation}\label{18}
\rho=\left(\frac{\alpha}{\beta}\right)\frac{b'}{r^2},
\end{equation}
\begin{equation}\label{19}
p= - \left(\frac{\alpha}{\beta}\right)\frac{b}{r^3},
\end{equation}
\begin{equation}\label{20}
p_t= \left(\frac{\alpha}{\beta}\right)\frac{b-b'r}{2r^3}.
\end{equation}
It is important to note that the field equations were modified solely in relation to Einstein's gravitational coupling constant, differing from GR. It is easy to show that
\begin{equation}\label{18a}
\rho(r)+p(r)=H(r)=\frac{\alpha}{\beta}\left(\frac{rb'-b}{r^2}\right),
\end{equation}
and
 \begin{equation}\label{18b}
\rho(r)+p_t(r)=H_1(r)=\frac{\alpha}{\beta}\left(\frac{rb'+b}{2r^2}\right),
\end{equation}
 It can be concluded that the outcome of this model is similar to $f(R, T)=\alpha R+ \beta T$, in the background of $f(R, T)$ theory, which was examined earlier in Ref. \cite{SR1}. As a result, it can be deduced that solutions that violate both the radial and lateral NEC in the framework of GR may still comply with the NEC when evaluated in the context of $f(R, L_m)$ gravity, as long as the condition
  \begin{equation}\label{18bb}
  \frac{\alpha}{\beta}<0,
 \end{equation}
 continues to hold. So, it is easy to show that utilizing an EoS in the form
\begin{equation}\label{24a}
p (r)=\omega\rho(r)
\end{equation}
in field equations leads to
\begin{equation}\label{241}
b(r)=r^m.
\end{equation}
where
\begin{equation}\label{27}
m=-\frac{1}{\omega}.
\end{equation}
The shape function (\ref{241}) is well-known in the realm of wormhole theory\cite{fq, fq2, SR1}. It meets all fundamental requirements necessary for the existence of traversable wormhole solutions. Consequently, it is crucial to take into account the asymptotically flat solution, which requires the condition that $m<1$ to be enforced. In order to achieve a positive energy density, the condition $m<0$ must be satisfied. Therefore, we can deduce that
\begin{equation}\label{271}
\omega> 0.
\end{equation}
As previously stated, the results of the field equations indicate that the linear model of $f(R,L_m)$ results in a modification of the gravitational coupling constant as outlined below.
\begin{equation}\label{27a}
G^f=\frac{\alpha}{\beta}\quad G^{GR},
\end{equation}
where $G^f$ is the gravitational coupling constant in the context of $f(R, L_m)$ and $G^{GR}$ is the gravitational coupling constant in the background of GR.
This finding demonstrates that the solutions discussed in \cite{fq, fq2, SR1} can be regarded as non-exotic solutions within the framework of $f(R,L_m)$ theory. As an example, we consider
\begin{equation}\label{f1}
b(r)=(\omega_\infty +D)^{1/\omega_\infty}(\omega_\infty r+D)^{-1/\omega_\infty}
\end{equation}
which was earlier discussed in \cite{SR1}, as a solution with variable EoS. It is easy to show that this shape function gives
\begin{equation}\label{f2}
\omega(r)=\frac{p(r)}{\rho(r)}=\omega_\infty+\frac{D}{r}
\end{equation}
in the context of $f(R, L_m)$ gravity. It is important to mention that  $\omega_\infty$ is referred  to the EoS parameter at large distance. Also, the EoS at the throat is
\begin{equation}\label{f3}
 \omega_0=\frac{p_0}{\rho_0}=\lim_{r\longrightarrow r_0}\omega(r) = \omega_\infty+D
\end{equation}
so
\begin{equation}\label{f4}
 D=\omega_0- \omega_\infty.
\end{equation}
According to \cite{SR1}, we can show that
\begin{equation}\label{f5}
 \rho_0=\lim_{r\longrightarrow r_0}\rho(r) = -\frac{\frac{\alpha}{\beta}}{\omega_0}
\end{equation}
is the energy density at the wormhole throat.
Using Eq. (\ref{f5}) shows that
\begin{equation}\label{f6}
 \omega_0=-\frac{\alpha}{\beta}(\rho_0)^{-1}.
\end{equation}
Using Eq.(\ref{f6}) indicates
\begin{equation}\label{f7}
p_0= \omega_0\rho_0=-\frac{\alpha}{\beta}.
\end{equation}
Certainly, this equation can be achieved by employing ( \ref{2}) and ( \ref{19}) for a general shape function.
One may use ( \ref{f7}) to define the connection between $ \frac{\alpha}{\beta}$ and radial pressure at the throat. As previously stated, the breach of the NEC is contingent upon the sign of $ \frac{\alpha}{\beta}$; thus, it can be inferred that the radial pressure value at the wormhole throat directly influences the violation of the ECs.

\subsection{ Nonlinear model }\label{subsec2}
A broader formulation, denoted as $f(R, L_m)=f_1(R)+f_2(R)f_3(L_m)$, has been suggested in the literature to tackle various curvature–matter interactions. This formulation illustrates the influence of curvature on the celestial system. The incorporation of $f_3(L_m)$ adds a dependency on the distribution of matter, thereby capturing the complex interplay between matter and curvature \cite{L4}. The function
\begin{equation}\label{28bb}
f(R, L_m)=\frac{R}{2}+L^\alpha_m
\end{equation}
is the next candidate to examine wormhole solutions. This model can be interpreted as an interaction between matter and gravitational fields through the metric tensor, which acts as a fundamental component in GR. This interaction aligns with the conventional perspective on how matter affects the curvature of spacetime. In situations where $\alpha = 1$, the model produces outcomes that correspond with the solutions recognized in GR. In this case, the field equations (\ref{3e}-\ref{3g}) give
\begin{equation}\label{29}
\rho=\left(\frac{b'}{ (2\alpha-1)r^2}\right)^\frac{1}{\alpha}.
\end{equation}
\begin{equation}\label{a29}
p=\frac{\rho\left( (\alpha-1)r^3 -b \rho^{-\alpha} \right)}{\alpha r^3}.
\end{equation}
\begin{equation}\label{25}
 p_t=\frac{\rho^{(1-\alpha)}\left( b-r b' +2(\alpha-1) r^3  \rho^\alpha \right)}{2\alpha r^3}.
\end{equation}
Numerous researchers have employed these field equations to derive wormhole solutions; however, none of these solutions fulfill the ECs. In this Section, we aim to identify solutions that comply with the ECs. We examine the shape function $b(r)=r^m$, which is among the most renowned in the literature. This shape function leads to
\begin{equation}\label{38}
\rho(r)=(\frac{m}{2\alpha-1})^{\frac{1}{\alpha}}r^{\frac{m-3}{\alpha}}.
\end{equation}
\begin{figure}
\subfloat (a){\includegraphics[width = 3in]{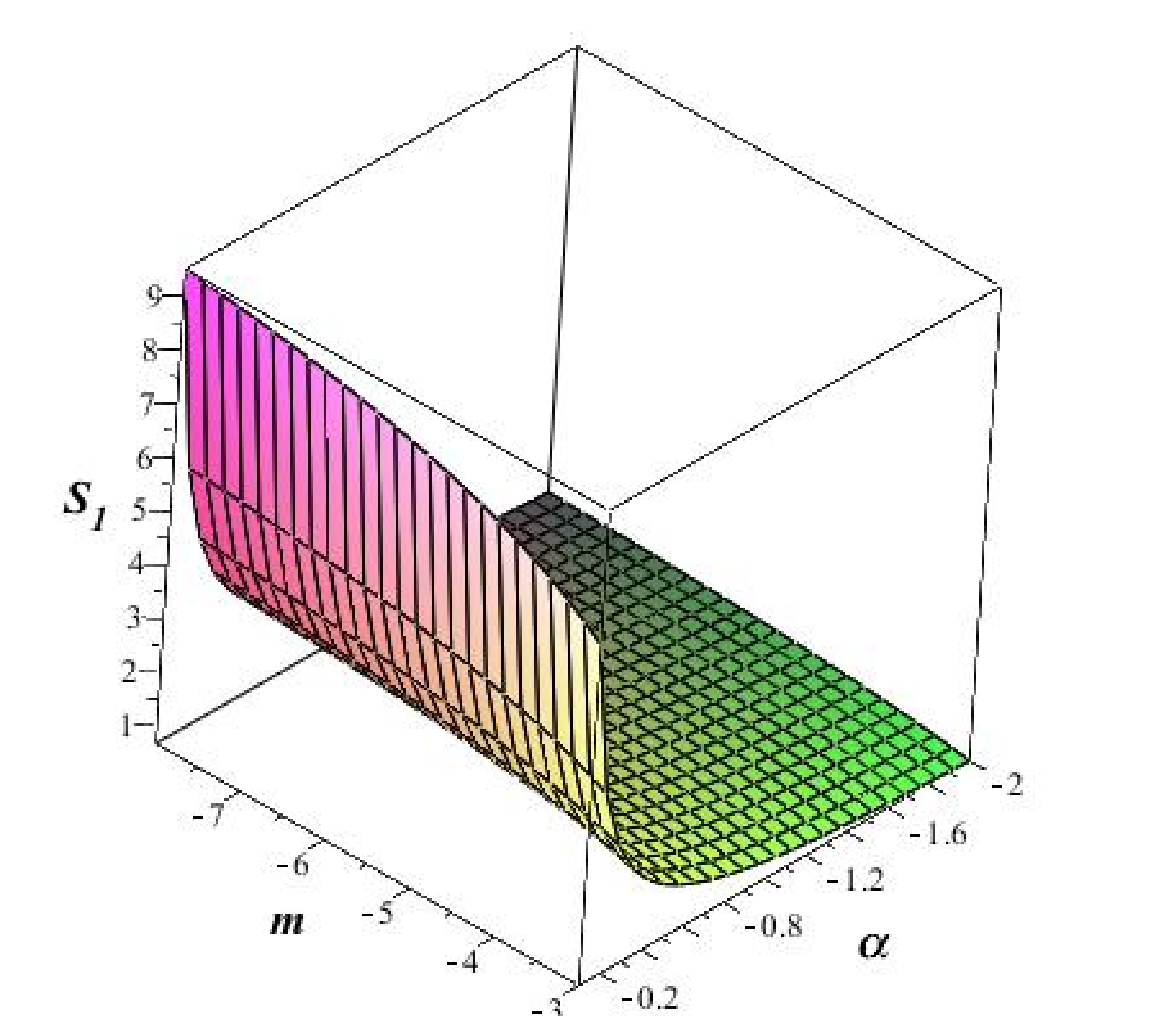}}\\
\subfloat(b){\includegraphics[width = 3in]{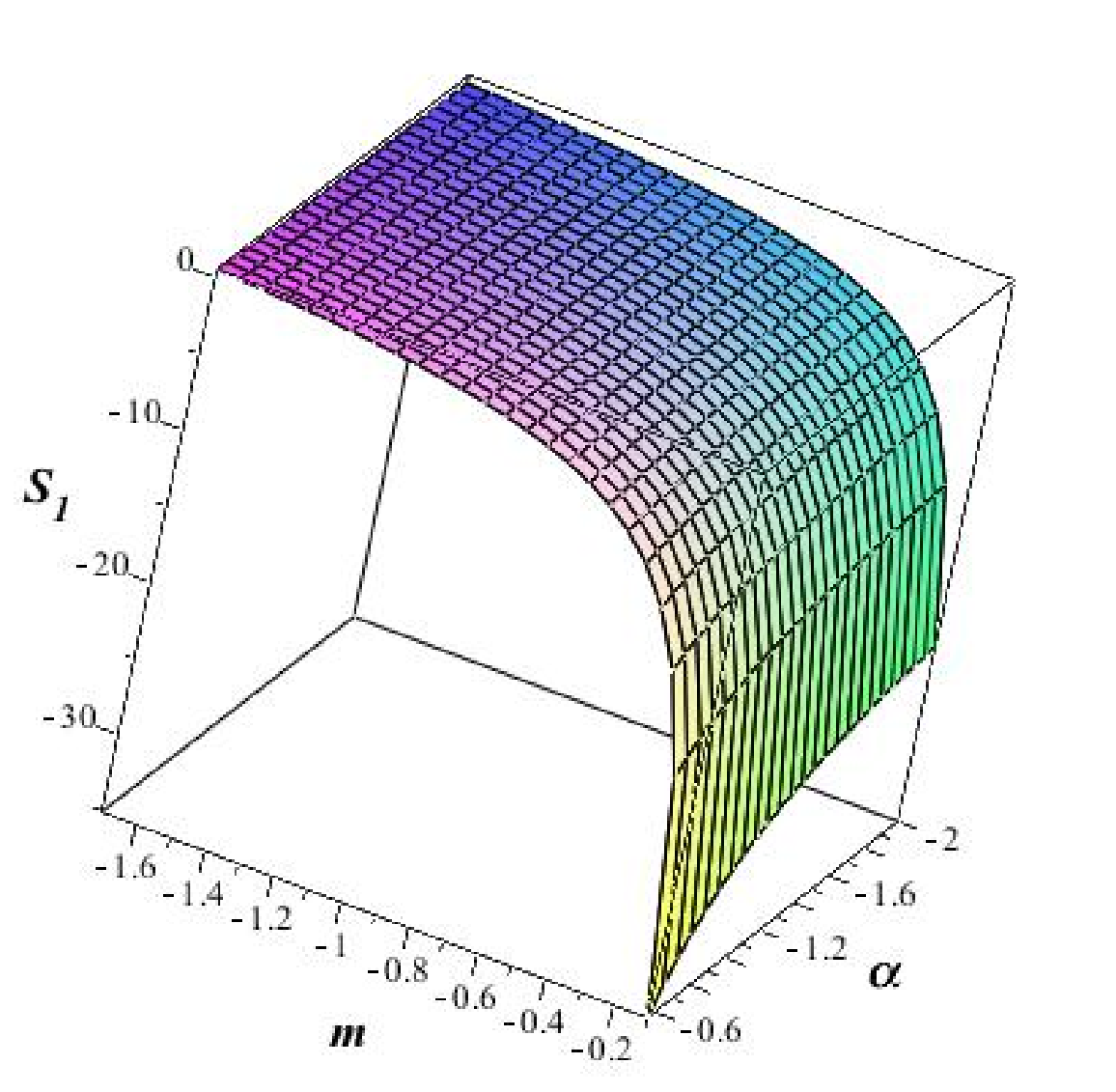}}
\caption{The graph depicts the correlation between $S_1(\alpha, m)$ and the variables $\alpha$ and $m$. It is clear that $S_1$ is positive within a certain range of $\alpha$ and $m$ (a), while $S_1$ is negative in a different range (b).}
\label{fig1}
\end{figure}

To have positive energy density, the condition
\begin{equation}\label{39}
\frac{m}{2\alpha-1}>0
\end{equation}
must hold. Since asymptotically flattens condition  requires $m<1$, the subsequent ranges are permissible for the parameters $m$ and $\alpha$
 \begin{equation}\label{39a}
\alpha>1/2, \qquad 0<m<1
\end{equation}
or
\begin{equation}\label{39b}
\alpha<1/2, \qquad m<0
\end{equation}
Using (\ref{29}) and (\ref{a29}) leads to
\begin{equation}\label{a22}
H(r)=\rho(r)+p(r)=S(\alpha,m)\,\rho(r),
\end{equation}
where
\begin{equation}\label{f30}
S(\alpha,m)=\left(2-\frac{1}{\alpha}\right) \left(1-\frac{1}{m}\right).
\end{equation}
We have taken into account a positive energy density, thus the condition $H>0$ holds while $S>0$ is also fulfilled. By using (\ref{39a}-\ref{f30}), it is easy to show that $H>0$ is reachable in the range
\begin{equation}\label{f31}
\alpha<0, \qquad m<0.
\end{equation}
or
\begin{equation}\label{39aa}
\alpha>1/2, \qquad 0<m<1
\end{equation}
The lateral NEC gives
\begin{equation}\label{8b1}
H_1(r)=\rho+p_t(r)=S_1(\alpha,m)\,\rho(r),
\end{equation}
where
\begin{equation}\label{209b}
S_1(\alpha,m)=\left(\frac{\alpha m+2\alpha-m-1}{\alpha m}\right).
\end{equation}
\begin{figure}
\centering
  \includegraphics[width=3 in]{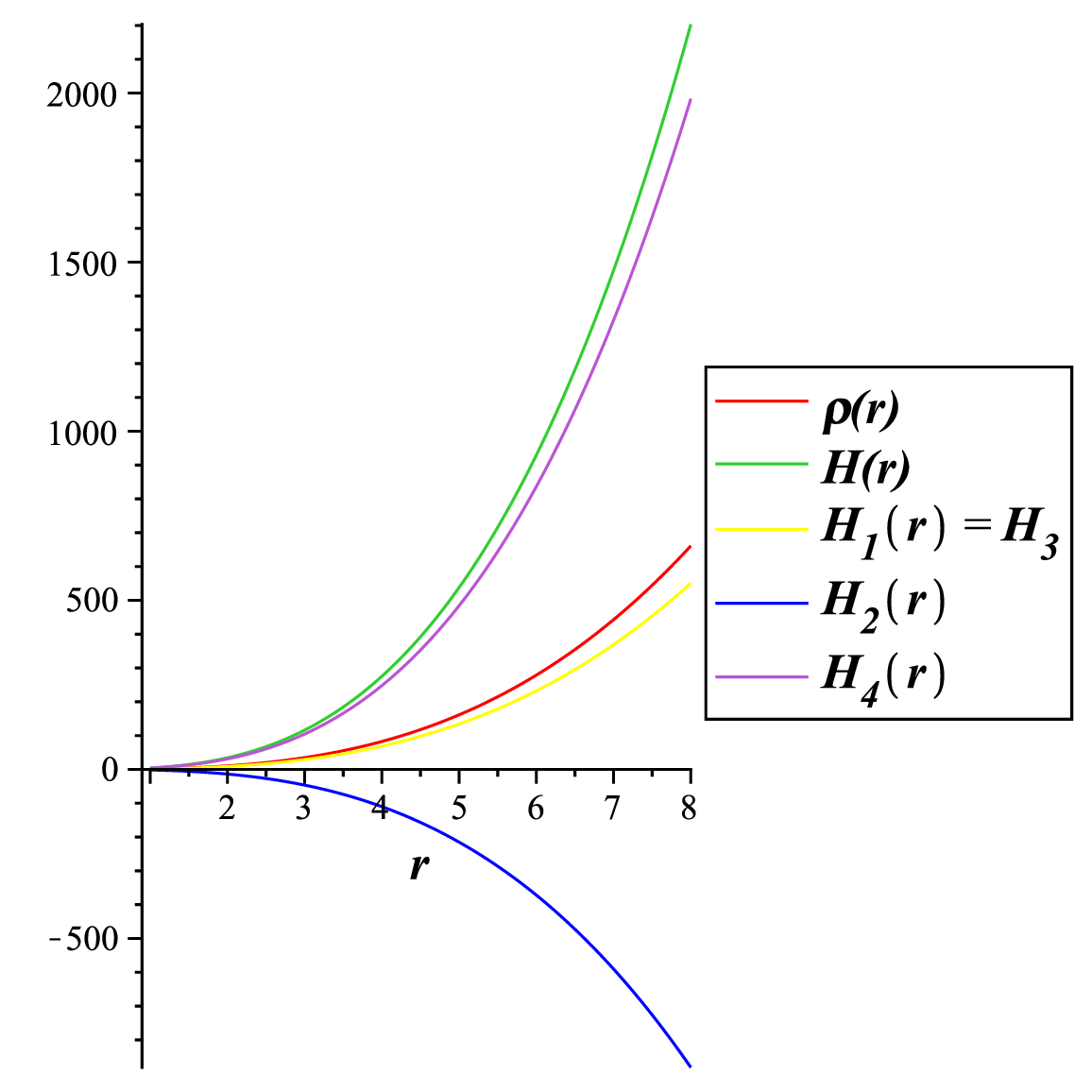}
\caption{The graph depicts the functions $ \rho(r)$(red), $H(r)$(green), $H_1(r)=H_3(r)$(yellow), $H_2(r)$(blue), and $H_4(r)$(violet) plotted against the radial coordinate for the shape function $b(r)=1/r^3$ within the framework of $f(R, L_m)=R/2+L_m^{-2}$. It is evident that all ECs, except for the DEC, are satisfied. See the text for details.}
 \label{fig2}
\end{figure}
We have plotted $S_1(\alpha,m)$ as a function of $\alpha$ and $m$ in Fig.(\ref{fig1}), which is positive for some range of  $\alpha$ and $m$. These results indicate that a power-law shape function can be considered as a non-exotic solution for some ranges of $\alpha$ and $m$. As an example, we set $alpha=-2$ and $m=-3$ which gives
\begin{equation}\label{s1}
b(r)=\frac{1}{r^3}.
\end{equation}
We have plotted $\rho(r), H(r), H_1(r), H_2(r), H_4(r)$ as a function of radial coordinate in Fig. (\ref{fig2}). This figure shows that the shape function $b(r)=1/r^3$ in the context of $f(R, l_m)=\frac{R}{2}+Lm^{-2}$ satisfies WEC, NEC, and SEC.

Let us investigate the EoS for $b(r)=r^m$, using (\ref{f30}) and (\ref{209b}) gives
\begin{equation}\label{s5}
\omega(\alpha,m)=\frac{p(r)}{\rho(r)}=S(\alpha,m)-1=1-\frac{m+2\alpha-1}{m \alpha}
\end{equation}
and
\begin{equation}\label{s6}
\omega_t(\alpha,m)=\frac{p_t(r)}{\rho(r)}=S_1(\alpha,m)-1=\frac{2\alpha-m-1}{2m \alpha}.
\end{equation}
Equations (\ref{s5}) and (\ref{s6}) explain that the power-law shape function admits linear EoS which is the most prevalent EoS in cosmology. It is evident that if one considers an EoS in the form of  $p=\omega \rho$ the resulted solution will be the power-law shape function.
It can be shown that $\omega$ approaches $\omega_t$ as $\alpha\rightarrow \frac{1}{2}$ however, such solutions  violate the ECs.

\section{Concluding remarks}

The traversable wormholes that display spherical symmetry have become a crucial focus in theoretical research, their feasibility depends on the existence of exotic matter. The fundamental aspect of the wormhole within the context of GR is the breach of ECs, and the exotic matter located at or near the throat is a substance that disregards the ECs. Modified gravitational theories, as opposed to GR, have demonstrated greater influence in the investigation of wormholes. We have analyzed a gravitational theory that incorporates a curvature–matter coupling, denoted by $f(R, L_m)$,  with some free parameters. The $f(R, L_m)$ represents an extension of the modified gravity $f(R)$ that includes a direct coupling of the arbitrary function of the Ricci curvature with the matter Lagrangian term.
 The incorporation of $L_m$ in the altered action facilitates a more complex interaction between gravity and matter, which may also influence the dynamics of astrophysical entities like galaxies and stars. The attractive features of $f(R, L_m)$ gravity, being a type of modified gravity, motivated us to explore wormholes in this context.
The assumption of a constant redshift function leads us to conclude that there are no variations in gravitational time dilation and that tidal forces are negligible. This not only ensures the safe transit of objects through the wormhole but also simplifies our study, enabling a clearer examination of the wormhole's properties.

 In this research, we have performed an extensive examination of wormhole solutions in the context of $f(R, L_m)$ gravity by considering two models for the function $f(R, L_m)$. The first model, $f(R, L_m)=\alpha R+\beta L_m$, as a linear model of $R$ and $L_m$, leads to a modified gravitational coupling constant. It is important to highlight that the parameters $\alpha$ and $\beta$ in this model are crucial for examining the breach of the ECs in wormholes.
 It has been indicated that the condition $\alpha/\beta<0$  may result in solutions that adhere to the ECs. In different terminology, it has been shown that solutions that breach both radial and lateral NEC at the same time, within the context of GR, can fulfill NEC in the background of linear $f(R, L_m)$ when a negative effective coupling constant is involved. It was shown that a linear EoS for radial pressure leads to a power-law shape function. So, the power-law shape function ($b(r)=r^m$) can be considered as a possible wormhole solution in the framework of $f(R, L_m)=\alpha R+\beta L_m$ which respects the ECs for $\alpha/\beta<0$ and $m<0$. Additionally, we highlighted that wormhole solutions presented within the framework of $f(R,T)$ \cite{SR1}and $f(t,T)$\cite{foad4}, which feature asymptotically linear EoS, can be regarded as non-exotic wormhole solutions in the context of $f(R, L_m)$.

In the second category of solutions, we have considered a $f(R, L_m)$ function in the form of (\ref{28bb}). This function has been extensively utilized by researchers. We have established the corresponding field equations, and by examining a power-law shape function ($b(r)=r^m$), we have identified the conditions on the free parameters $m$ and $\alpha$ necessary for obtaining non-exotic wormhole solutions. Our research indicates that within the interval $\alpha<0$ and a specfic range of $m$, the WEC, NEC, and SEC are fulfilled, whereas the DEC is not satisfied. It has been shown that the power-law shape function allows for a Linear EoS, or conversely, assuming a linear EoS for radial pressure results in a power-law shape function. It has been demonstrated that approximately isotropic solutions can be attained within the framework of $f(R, L_m)$; however, this class of solutions violates the ECs.
The various forms of $f(R, L_m)$ represent distinct physical phenomena and can result in a variety of implications for cosmology and astrophysics. Linear functions are typically more straightforward and easier to analyze, whereas power-law forms can yield more intricate dynamics. Power-law and generalized polynomial functions offer enhanced flexibility in modeling the effects of modified gravity, rendering them appropriate for fitting observational data. Each selection of $f(R, L_m)$ can produce different predictions concerning the Universe's behavior, making the investigation of these functions a fertile ground for theoretical inquiry. Our research indicates that non-exotic matter can be accessed within the framework of two different forms of the $f(R, L_m)$ functions considered in this study. However, the linear form of $f(R, L_m)$ presents a broad spectrum of solutions.

In summary, for a revised cosmological theory to be considered credible, it must undergo thorough investigations that include detailed comparisons with key observational data like Baryon Acoustic Oscillations (BAO), the Cosmic Microwave Background (CMB), and the constraints from Big Bang Nucleosynthesis (BBN). This process is essential for validating the theory and determining its viability in explaining the universe's natural phenomena. This research elucidates that the $f(R, L_m)$ theory has the potential to remove the need for exotic matter within the framework of wormhole theory. Following the derivation of spherically symmetric solutions, it would be beneficial to evaluate $f(R,L_m)$ gravity in relation to data acquired from the Solar System. Consequently, researchers have extensively explored various aspects of the cosmos within the framework of $f(R,L_m)$ gravity in recent years.

\end{document}